\documentclass[preprintnumbers,amsmath,amssymb,floatfix,11pt,prd,onecolumn,superscriptaddress,nofootinbib]{revtex4}
\usepackage{hyperref}
\usepackage{footnote}
\usepackage{amssymb}
\usepackage{amsmath}
\usepackage{graphicx}
\usepackage{dcolumn}
\usepackage{bm}
\usepackage{color}
\usepackage{xcolor}
\usepackage{float}
\usepackage{lipsum}
\allowdisplaybreaks
\usepackage{subfig}
\usepackage{comment}
\usepackage{multirow}
\usepackage[figurename=FIGURE,tablename=TABLE]{caption}
\usepackage{threeparttable}
\usepackage{hyperref}
\hypersetup{
    colorlinks=true, 
    linktoc=all,     
    linkcolor=magenta,
    citecolor=red
}

\newcommand{\ba}{\begin{eqnarray}}
\newcommand{\ea}{\end{eqnarray}}

\begin{document}

\title{Constraining non-minimally coupled squared-Quartic Hilltop Inflation in light of ACT observations}

\author{Jureeporn Yuennan}
\email{jureeporn\_yue@nstru.ac.th}
\affiliation{Faculty of Science and Technology, Nakhon Si Thammarat Rajabhat University, Nakhon Si Thammarat, 80280, Thailand}

\author{Farruh Atamurotov}
\email{atamurotov@yahoo.com}
\affiliation{Urgench State University, Kh. Alimdjan str. 14, Urgench 220100, Uzbekistan}

\author{Phongpichit Channuie}
\email{phongpichit.ch@mail.wu.ac.th}
\affiliation{School of Science \& College of Graduate Studies, Walailak University, Nakhon Si Thammarat, 80160, Thailand}

\date{\today}

\begin{abstract}
The combination of the data from the Dark Energy Spectroscopic Instrument (DESI) with the recent measurements from the Atacama Cosmology Telescope (ACT) indicate that the scalar spectral index \( n_s \) has a larger value than the Planck 2018 which leads to tension within standard inflationary models. In this study in order to explain the new data,  We consider the squared-Quartic Hilltop inflation potential \( V(\phi) = V_0 [1 - \lambda (\phi/M_p)^4]^2 \) within the Einstein and Jordan frames. In the Jordan frame we introduce the coupling term \( \xi \phi^2 R \) and we calculate analytic expressions for the slow-roll parameters, scalar spectral index, and tensor-to-scalar ratio on the weak and strong coupling regimes. In the weak limit (\( \xi \ll 1 \)), perturbative corrections slightly increase \( n_s \) and suppress \( r \), leading to \( n_s \simeq 0.9743 \) and \( r \sim 7.8 \times 10^{-5} \) for representative parameters \( \lambda = 10^{-3}, \xi = 10^{-3}, {\cal N} = 117 \), values which are in agreement with the joint Planck--ACT--DESI (P-ACT-LB) constraints. On the other hand, for a strong coupled (\( \xi \gg 1 \)), the conformal rescaling provides an exponentially flat potential plateau, which allows us to calculate \( n_s \approx 0.9743 \) with \( r \lesssim 5 \times 10^{-4} \) for \( {\cal N} = 65{-}70 \), consistent with ACT and BK18 bounds. The associated energy scale of inflation, \( V_0^{1/4} \sim 10^{-3}{-}10^{-2} M_p \), remains compatible with high-scale inflationary scenarios. 
\end{abstract}


\maketitle

\newpage
\section{Introduction}
Recent analyses that combine observations from the Atacama Cosmology Telescope (ACT)~\cite{ACT:2025fju,ACT:2025tim} and the Dark Energy Spectroscopic Instrument (DESI) survey~\cite{DESI:2024uvr,DESI:2024mwx} have led cosmologists to reexamine the standard paradigm of inflation. The latest ACT measurements reveal a $\sim2\sigma$ deviation in the scalar spectral index of primordial curvature perturbations compared to the {\it Planck} 2018 results~\cite{Planck:2018jri}, which indicate that the conventional inflationary framework may should revised. Inflation remains the simplest mechanism to solve cosmological problems as the flatness, horizon, and monopole problems. Furthermore, it provides a description for the origin of primordial fluctuations that seeded the large-scale structures in the Universe and appear today as temperature anisotropies in the cosmic microwave background (CMB)~\cite{Starobinsky:1980te,Sato:1981qmu,Guth:1980zm,Linde:1981mu,Albrecht:1982wi}.

These primordial perturbations are mainly described by two main observables, the scalar spectral index, $n_s$, quantifying the scale dependence of scalar modes, and the tensor-to-scalar ratio, $r$, which measures the amplitude of primordial gravitational waves relative to scalar perturbations. In standard slow-roll inflation, both parameters are functions of the number of $e$-folds, $N$, between horizon exit and the end of inflation, allowing direct comparison with observations.
A widely known feature shared by many inflationary models is the “universal attractor” prediction, $n_s = 1 - \tfrac{2}{N}$, which arises in various frameworks such as $\alpha$-attractors~\cite{Kallosh:2013tua,Kallosh:2013hoa,Kallosh:2013maa,Kallosh:2013yoa,Kallosh:2014rga,Kallosh:2015lwa,Roest:2015qya,Linde:2016uec,Terada:2016nqg,Ueno:2016dim,Odintsov:2016vzz,Akrami:2017cir,Dimopoulos:2017zvq,Barrow:2016qkh}, the Starobinsky $R^2$ model~\cite{Starobinsky:1980te}, and Higgs inflation with strong non-minimal coupling to gravity~\cite{Kaiser:1994vs,Bezrukov:2007ep,Bezrukov:2008ej}. Finite-temperature one-loop corrections to Higgs inflation were explored in Ref.~\cite{Chang:2019ebx}, showing that thermal effects can improve consistency with {\it Planck} data. Comparable results have also been reported in composite inflaton models~\cite{Karwan:2013iph,Channuie:2012bv,Bezrukov:2011mv,Channuie:2011rq}, as discussed in Refs.~\cite{Channuie:2014ysa,Samart:2022pza}.
For a typical choice of $N = 60$, the universal attractor yields $n_s \approx 0.9667$, in excellent agreement with the {\it Planck} 2018 value of $n_s = 0.9649 \pm 0.0042$~\cite{Planck:2018jri}. Nevertheless, recent ACT measurements~\cite{ACT:2025fju,ACT:2025tim} suggest a slightly larger spectral index. The joint {\it Planck}–ACT (P-ACT) analysis finds $n_s = 0.9709 \pm 0.0038$, while the inclusion of CMB lensing and baryon acoustic oscillation (BAO) data from DESI (P-ACT-LB) increases this to $n_s = 0.9743 \pm 0.0034$. These updated results create notable tension with the universal attractor framework, excluding many of its realizations at about the $2\sigma$ level and even challenging the predictions of the Starobinsky $R^2$ model~\cite{ACT:2025fju}.
In light of this, Ref.~\cite{Gialamas:2025kef} revisited Higgs-like inflation incorporating radiative corrections to assess its compatibility with ACT data. This unexpected deviation has stimulated numerous theoretical efforts aimed at reconciling inflationary scenarios with the new ACT constraints~\cite{Kallosh:2025rni,Gao:2025onc,Liu:2025qca,Yogesh:2025wak,Yi:2025dms,Peng:2025bws,Yin:2025rrs,Byrnes:2025kit,Wolf:2025ecy,Aoki:2025wld,Gao:2025viy,Zahoor:2025nuq,Ferreira:2025lrd,Mohammadi:2025gbu,Choudhury:2025vso,Odintsov:2025eiv,Odintsov:2025wai,Q:2025ycf,Zhu:2025twm,Kouniatalis:2025orn,Hai:2025wvs,Dioguardi:2025vci,Yuennan:2025kde,Oikonomou:2025xms,Oikonomou:2025htz,Odintsov:2025jky,Aoki:2025ywt,Gialamas:2025kef,Gialamas:2025ofz,Yuennan:2025tyx,Pallis:2025nrv,Pallis:2025gii,Yuennan:2025inm,Lynker:2025wyc}, with a comprehensive synthesis of these developments presented in Ref.~\cite{Kallosh:2025ijd}.

In the present study, we examine both minimally and non-minimally coupled inflationary scenarios, defined in the Einstein and Jodan frames respectively, in which the early accelerated expansion of the Universe is driven by a class of squared-Quartic Hilltop Inflation \cite{Hoffmann:2021vty}, see also Refs.\cite{Dimopoulos:2020kol,German:2020rpn}.

In quantum field theory formulated on a curved background, it is well established that a coupling between the spacetime curvature and the scalar field is \emph{unavoidable}. Even if one begins with a minimally coupled scalar Lagrangian, quantum corrections at one-loop order inevitably generate a term proportional to $R\phi^{2}$ in the effective action. This interaction, commonly written as $\tfrac{1}{2}\xi R\phi^{2}$, represents the simplest generally covariant extension of the flat-space theory and is required to ensure the renormalizability of scalar fields in curved spacetime. In the absence of such a term, the divergences arising from loop corrections cannot be absorbed into redefinitions of existing parameters, rendering the theory inconsistent beyond tree level.

The emergence of this non-minimal coupling can be rigorously demonstrated using the DeWitt-Schwinger (heat-kernel) expansion for the one-loop effective action \cite{Schwinger:1951nm,DeWitt:1975ys}. The corresponding Seeley-DeWitt coefficient $a_{2}$ contains cross terms of the form $(\xi - \tfrac{1}{6})\,R\,V''(\phi)$, which give rise to divergences proportional to $R\phi^{2}$. Consequently, the inclusion of $\xi R\phi^{2}$ in the bare action is mandatory to cancel these divergences through renormalization. The special value $\xi = 1/6$ corresponds to the conformal coupling, for which the classical massless scalar action becomes invariant under conformal transformations in four dimensions and the $R\phi^{2}$ divergence vanishes. Therefore, from both the renormalization and symmetry standpoints, the non-minimal coupling term is a natural and inevitable feature of any consistent scalar field theory in curved spacetime.

Historically, the necessity of this coupling was recognized in the pioneering work of Callan, Coleman, and Jackiw~\cite{Callan:1970ze}, who first identified the conformal value $\xi = 1/6$ in flat space. This was later formalized in the comprehensive analyses of Refs.\cite{Buchbinder:1985ew,Odintsov:1990mt,Muta:1991mw,Elizalde:1994im,Buchbinder:1989bt,Chernikov:1968zm}, and Parker and Toms~\cite{Parker:1983pe}. In modern applications, such as inflationary cosmology, scalar–tensor gravity, and modified gravity theories, the same coupling plays a central role in connecting matter fields to the gravitational sector, demonstrating how quantum consistency conditions at the microscopic level naturally extend to macroscopic phenomena.

The structure of the paper is as follows. Section~(\ref{II}) provides an overview of scalar–gravity coupling theories and presents the formalism of the non-minimally coupled framework of squared- Quartic Hilltop Inflation. It also derives the analytic expressons for the slow-roll parameters and the inflationary observables $n_s$ and $r$ for both minimal and non-minimal couplings. Section~(\ref{III}) compares the model predictions with the most recent ACT and DESI observations. The main results and conclusions are summarized in Section~\ref{Con}.

\section{Slow-roll Analysis}\label{II}
We revisit the inflationary framework in which a scalar field is non-minimally coupled to gravity. The dynamics follow from a general gravitational action that includes a non-minimal coupling between the scalar field and the Ricci scalar, together with an arbitrary effective potential $V(\phi)$, given by
\begin{eqnarray}
S = \int d^4x \sqrt{-g} 
\left[
\frac{\Omega^2(\phi)}{2\kappa^2} R
- \frac{1}{2} g^{\mu\nu} \partial_\mu \phi \partial_\nu \phi
- V(\phi)
\right],
\label{1}
\end{eqnarray}
where $R$ denotes the Ricci scalar and $\Omega^2(\phi) \equiv 1 + \kappa^2 \xi \phi^2$. Here, $\xi$ quantifies the non-minimal coupling between the scalar field and gravity, $\kappa^2 = 8\pi G = M_p^{-2}$ with $M_p$ being the reduced Planck mass, and $V(\phi)$ represents the potential energy of the inflaton.  
Evidently, in the limit $\xi = 0$, the theory reduces to the standard minimally coupled case. Because of the coupling term $\Omega^2(\phi)$, the field equations are more complicated. To cast them into the canonical Einstein form, we apply a conformal transformation of the metric,
\begin{gather}
g_{\mu\nu} \rightarrow \hat{g}_{\mu\nu} = \Omega^2 g_{\mu\nu}.
\label{2}
\end{gather}
Under this transformation, the Ricci scalar and other geometric quantities are redefined, leading to the Einstein-frame action
\begin{eqnarray}
S = \int d^4x \sqrt{-g}
\left[
\frac{1}{2\kappa^2} R
- \frac{1}{2} F^2(\phi) \hat{g}^{\mu\nu} \partial_\mu \phi \partial_\nu \phi
- U(\phi)
\right],
\label{3}
\end{eqnarray}
where
\begin{eqnarray}
F^2(\phi) &=& \frac{1 + \kappa^2 \xi \phi^2 (1 + 6\xi)}{(1 + \kappa^2 \xi \phi^2)^2},
\label{4}\\
U(\phi) &=& \frac{V(\phi)}{(1 + \kappa^2 \xi \phi^2)^2}.
\label{5}
\end{eqnarray}
After eliminating the non-minimal coupling to gravity, the scalar field $\phi$ acquires a non-canonical kinetic term. This can be recast in canonical form by defining a new field $\chi(\phi)$, related to $\phi$ through:
\begin{eqnarray}
\frac{1}{2}{\tilde g}^{\mu\nu}\partial_{\mu}\chi(\phi)\partial_{\nu}\chi(\phi) = \frac{1}{2}\Bigg(\frac{d\chi}{d\phi}\Bigg)^{2}{\tilde g}^{\mu\nu}\partial_{\mu}\phi\partial_{\nu}\phi\,,
\label{611}
\end{eqnarray}
where $d\chi/d\phi$ is given by
\begin{eqnarray}
\frac{d\chi}{d\phi} = 
\sqrt{\frac{1 + \kappa^2 \xi \phi^2 (1 + 6\xi)}{(1 + \kappa^2 \xi \phi^2)^2}}\,.
\label{6}
\end{eqnarray}
Having rewritten the expression in the canonically normalized field, the action now reads:
\begin{eqnarray}
S = \int d^4x \sqrt{-g}
\left[
\frac{1}{2\kappa^2} R
- \frac{1}{2} g^{\mu\nu} \partial_\mu \chi \partial_\nu \chi
- U(\chi)
\right].
\label{7}
\end{eqnarray}
Assuming slow-roll dynamics, the inflaton evolves gradually along the potential, and the first two slow-roll parameters are defined in terms of $U$ and $\chi$ as
\begin{eqnarray}
\epsilon = \frac{M_p^2}{2} 
\left( \frac{U_{\chi}}{U} \right)^2,
\qquad
\eta = M_p^2 \left(
\frac{U_{\chi\chi}}{U}\right),
\label{8}
\end{eqnarray}
We will, however, express everything in terms of $\phi$, such that we don’t need to solve for an explicit solution of (\ref{6}). We therefore obtain
\begin{eqnarray}
\epsilon = \frac{M_p^2}{2} 
\left( \frac{V_{\phi}}{\chi_{\phi} V } \right)^2,
\qquad
\eta = M_p^2 \left(
\frac{V_{\phi\phi}}{\chi_{\phi}^2 V }
- \frac{ \chi_{\phi\phi}V_{\phi}}{\chi_{\phi}^3 V }
\right),
\label{81}
\end{eqnarray}
Inflation persists as long as $\epsilon, |\eta| \ll 1$, and it ends when one of them reaches unity, typically $\epsilon(\phi_{\text{end}}) = 1$.  The key observable quantities of interest are the scalar spectral index $n_s$, which measures the scale dependence of curvature perturbations, and the tensor-to-scalar ratio $r$, quantifying the amplitude of primordial tensor modes:
\begin{eqnarray}
n_s = 1 - 6\epsilon + 2\eta, 
\qquad 
r = 16\epsilon.
\label{9}
\end{eqnarray}
The amplitude of scalar perturbations at the pivot scale $k_\star$ is
\begin{gather}
P_{R\star} = \frac{1}{24\pi^2 M_p^4}
\frac{V(\phi_\star)}{\epsilon(\phi_\star)} \Big|_{k = k_\star},
\label{10}
\end{gather}
where $\phi_\star$ is the field value at horizon exit.  
We use the value \(P_R = 2.0933\times 10^{-9}\) at the pivot scale 
\(k_\star = 0.05~\text{Mpc}^{-1}\), in agreement with the Planck~2018 measurements. These data also provide constraints on the scalar spectral index, \(n_s = 0.9649 \pm 0.0042\) (68\% C.L.), and set an upper bound on the tensor-to-scalar ratio, \(r < 0.056\), based on the combined Planck and BICEP2/Keck Array results. We now compare the predictions of the minimally and non-minimally coupled inflationary models.

\subsection{Minimally coupled Scenario revisited}
There are various of studies which investigate inflation with the use of of hilltop (HI) and hilltop-squared (HSI) potentials, see for instance  \cite{Boubekeur:2005zm,Kinney:1995cc,German:2020rpn,Hoffmann:2021vty,Lillepalu:2022knx,Lynker:2023hfm} and references therein. In this study we consder the scalar field potential function
\begin{gather}
V(\phi) = V_{0} \left[1-\lambda  \left(\frac{\phi }{M_{p}}\right)^4\right]^2,
\label{12}
\end{gather}
which was later motivated within the braneworld context in Ref.~\cite{Kallosh:2019jnl}. We introduce the dimensionless variable $\psi \equiv \phi/M_{p}$, where now the slow-roll parameters read
\begin{gather}
\epsilon = \frac{32 \lambda^2 \psi^6}{\left(1-\lambda  \,\psi^4\right)^2},
\qquad
\eta = \frac{32 \lambda^2 \psi^6}{\left(1-\lambda \psi^4\right)^2}-\frac{24 \lambda \psi^2}{\left(1-\lambda \psi^4\right)}.
\label{13}
\end{gather}
The value of the scalar field at the end of inflation, $z_{\text{end}}$, is determined by the condition $\epsilon = 1$, yielding
\begin{gather}
\frac{32 \lambda^2 \psi_{\rm end}^6}{\left(1-\lambda  \,\psi_{\rm end}^4\right)^2} = 1 \longrightarrow  \begin{cases}
      \psi^{2}_{\text{end}}\simeq \frac{1}{\sqrt{\lambda}}, & \text{if}\ \lambda \ll 1 \\ \psi^{3}_{\text{end}}\simeq
      \frac{1}{4 \sqrt{2} \lambda }, & \text{if}\ \lambda \gg 1 
    \end{cases},
\label{14}
\end{gather}
where only the first solution corresponds to a physically viable solution when $\lambda\ll 1$. 
Furthermore, the number of $e$-folds is defined as
\begin{gather}
N = \frac{1}{M_{p}^{2}} \int_{\phi_{\text{end}}}^{\phi_{\star}} \frac{\chi_{\phi} V }{V_{\phi}}\,d\phi\quad\longrightarrow\quad {\cal N}=N+N_{\rm end}=\frac{1}{16 \lambda  \psi^{2}}+\frac{\psi^{2}}{16},
\end{gather}
Thus, the scalar field at the horizon crossing has the value
\begin{gather}
\psi^{2}_{\star} = 8\,{\cal N}\,\Sigma\quad{\rm with}\quad \Sigma =1-\sqrt{1-\frac{1}{\Pi }},\quad {\rm and}\quad\Pi =64 \lambda {\cal N}^2\,,
\label{15}
\end{gather}
in which we have considered
\begin{eqnarray}
N_{\rm end}=\frac{1}{16 \lambda  \psi^{2}_{\rm end}}+\frac{\psi^{2}_{\rm end}}{16}=\frac{1}{8 \sqrt{\lambda }}\,\,\,{\rm for}\,\,\,\lambda \ll 1\,,
\label{15}
\end{eqnarray}
and $\psi^{2}_{\rm end}$ is defined by Eq.(\ref{14}). Therefore, the spectral indices $n_{s}$ and $r$ are expressed 
\begin{eqnarray}
n_{s} &=& 1 - 6\epsilon + 2\eta 
\nonumber\\
&=& 1+16 \lambda  \psi^2 \left(\frac{4 \lambda  \psi^4}{\left(1-\lambda \psi^4\right)^2}-\frac{3}{1-\lambda  \psi^4}-\frac{12 \lambda  \psi}{\left(1-\lambda \psi^2\right)^2}\right), 
\\
r &=& 16\epsilon = \frac{512 \lambda^2 \psi^6}{\left(1-\lambda \,\psi^4\right)^2},
\label{16}
\end{eqnarray}
or equivalently, in terms of the number of $e$-folds
\begin{eqnarray}
n_{s} &=& 1-\frac{6 \Pi \, \Sigma }{{\cal N} \left(1-\Pi \,\Sigma^2\right)}-\frac{16 \Pi^2 \,\Sigma^3}{{\cal N} \left(1-\Pi \, \Sigma ^2\right)^2},
\label{ns1}\\
r &=& \frac{64 \Pi^2 \,\Sigma ^3}{{\cal N}\left(1-\Pi  \Sigma ^2\right)^2}.
\label{17}
\end{eqnarray}


\subsection{Non-minimally coupled Scenario}
We continue our study with the analysis the non-minimally coupled model, where the inflaton field is coupled to the gravitational field. We employ a conformal transformation to write the model in the Einstein frame, where now the effective scalar field potential is derived
\begin{eqnarray}
U(\chi(\phi)) \equiv \frac{V(\phi(\chi))}{\left(1 + \kappa^{2}\xi \phi^{2}(\chi)\right)^{2}},
\label{pot}
\end{eqnarray}
where $\phi(\chi)$ is defined by Eq.~(\ref{6}). From the form of the scalar field potentia (\ref{pot}), we conclude that the coupling function affects the dynamics in the Einstein frame. 
\subsubsection{Small Field Approximation}
Let us now consider the regime in which the quantity $\varphi \equiv \sqrt{\xi}\phi/M_{p}$ is much smaller than unity, i.e., $\varphi \ll 1$. 
This assumption corresponds to the weak-coupling limit, 
where the contribution from the non-minimal coupling term becomes subdominant. 
In this limit, the field redefinition between the Jordan and Einstein frames, 
given by Eq.~(\ref{6}), can be greatly simplified. 
Here, one obtains for this case
\begin{eqnarray}
\chi_{\phi} \equiv \frac{d\chi}{d\phi}
= \Bigg(\frac{(6\xi + 1) \varphi^2 + 1}{(\varphi^2 + 1)^2}\Bigg)^{1/2} \simeq 1\,.
\label{6zz}
\end{eqnarray}
Hence, to a good approximation, we find that $\chi \simeq \phi$, 
which implies that the canonical field in the Einstein frame nearly coincides with the original scalar field in the Jordan frame. 
This result allows one to treat the non-minimally coupled model effectively as a minimally coupled one in the small-field limit, 
significantly simplifying the dynamical.

\subsubsection{Large Field Approximation}
We now proceed to analyze the large-field approximation, where the behavior of the model becomes significantly more interesting and physically relevant for inflationary dynamics. 
In this regime, the presence of the coupling $\xi$ substantially modifies both the shape of the potential and the kinetic structure of the theory in the Einstein frame, leading to a rich phenomenology. 

To capture the essential features of this regime, we consider two distinct cases depending on the strength of the coupling parameter $\xi$. 
The first corresponds to the weakly coupled case ($\xi \ll 1$), 
where deviations from the minimally coupled scenario appear only as small perturbative corrections. The second corresponds to the strongly coupled case ($\xi \gg 1$), 
where the conformal rescaling strongly suppresses the potential at large field values, typically giving rise to an exponentially flat plateau. Such flattening of the potential is a characteristic feature of many successful inflationary models, 
including Higgs inflation and $\alpha$-attractor–type scenarios, and it plays a crucial role in ensuring slow-roll conditions consistent with CMB observations. 

\subsubsection*{2.1) Weakly coupled case ($\xi \ll 1$)}
Let us now consider the behavior of the field redefinition in the presence of a small but non-zero 
non-minimal coupling parameter $\xi$. 
Expanding the corresponding expression for $\chi_{\phi} \equiv d\chi/d\phi$ to first order in $\xi$, 
we obtain
\begin{eqnarray}
\chi_{\phi} \equiv \frac{d\chi}{d\phi}
= \Bigg(\frac{(6\xi + 1) \varphi^2 + 1}{(\varphi^2 + 1)^2}\Bigg)^{1/2} 
\simeq 1 - \frac{\psi^2}{2}\,\xi + {\cal O}\!\left(\xi^{2}\right)\,,
\label{6zz}
\end{eqnarray}
where $\sigma \equiv \phi/M_{p}$ and $\psi \equiv \sqrt{\xi}\phi/M_{p} = \sqrt{\xi}\,\sigma$. 
This perturbative expansion indicates that for small $\xi$, 
the canonical normalization of the field $\chi$ deviates only slightly from the minimally coupled case. 
The correction term proportional to $\xi\, \sigma^{2}$ introduces a mild suppression to the canonical field, 
reflecting the influence of the non-minimal coupling on the kinetic structure of the theory.

At leading order, integration of expression Eq.~(\ref{6zz}) leads to the asymptotic behaviour
\begin{eqnarray}
\chi \simeq \phi - \frac{\xi}{6}\frac{\phi^{3}}{M_{p}^{2}} + {\cal O}\!\left(\xi^{2}\right)\,,
\label{chiexpand}
\end{eqnarray}
which shows that $\chi$ and $\phi$ remain approximately equivalent for small field amplitudes, 
with a cubic-order correction due to the coupling. 
This deviation becomes increasingly important as $\phi$ grows, 
signaling the onset of nonlinear effects associated with the gravitational coupling term $\xi \phi^{2} R$. 
Such small corrections play a pivotal role in determining the shape of the Einstein-frame potential $U(\chi)$, 
and consequently in refining the predictions for inflationary observables such as the spectral index $n_{s}$ and tensor-to-scalar ratio $r$ in the weak-coupling regime. 

Therefore, the slow-roll parameters take the form
\begin{gather}
\epsilon \simeq \frac{\left(32 \lambda ^2 \sigma^6\right)}{\left(1-\lambda  \sigma^4\right)^2}\left(1+\xi \sigma^2\right),
\qquad
\eta \simeq -\frac{8 \lambda  \xi \sigma^4}{1-\lambda \sigma^4}+\frac{\left(32 \lambda^2 \sigma^6-24 \lambda \sigma^2 \left(1-\lambda \sigma^4\right)\right)}{\left(1-\lambda \sigma^4\right)^2}\left(1+\xi \sigma^2\right) .
\label{13n}
\end{gather}
We observe that in the limit $\xi=0$ the expressions of $\epsilon$ and $\eta$ for the minimally
coupled case Eq.(\ref{13}) are simply recovered. Moreover, the number of e-folds between the horizon exit of a given scale and the end of inflation, $N$, are calculated by the expression
\begin{eqnarray}
N=\frac{1}{M^{2}_{p}}\int^{\phi_{\star}}_{\phi_{\rm end}}\frac{V_{\phi}}{\chi_{\phi} V}d\phi=\frac{1}{16 \lambda \sigma^2}+\frac{\sigma^2}{16}-\Bigg(\frac{\sigma^4}{64}+\frac{\log (\sigma)}{16 \lambda }\Bigg)\xi\Bigg|^{\sigma_{\star}}_{\sigma_{\rm end}}+{\cal O}(\xi^{2})\,.\label{No}
\end{eqnarray}
However, it is more convenient to explore the small regime $\xi$, as this allows one to verify the consistency of the results with the minimally coupled limit, $\xi = 0$. This assumption also facilitates the application of the perturbative approach. Using a perturbation method for (\ref{13n}) in the small parameters $\xi$ and searching
for a solution to first order in $\xi$ to this condition of the type:
\begin{eqnarray}
\sigma_{\rm end}=\sigma_{0}+\sigma_{1}\xi, \quad{\rm with}\quad \sigma_{0}\simeq (1/\lambda)^{1/4}\,,
\end{eqnarray}
we end up with the solution for $\epsilon(\phi_{\rm end})=1$:
\begin{eqnarray}
\sigma_{\rm end}=z_{0}-\frac{\sigma^3_{0} \left(1-\lambda \sigma^4_{0}\right)}{2 \lambda \sigma^4_{0}+6}\xi+{\cal O}(\xi^{2})=\sigma_{0}.
\end{eqnarray}
We remark that $\sigma_{\rm end}$ does not depend on $\xi$. From Eq.(\ref{No}), we find up to the 1st-order in $\xi$:
\begin{eqnarray}
N=\frac{1}{M^{2}_{p}}\int^{\phi_{\star}}_{\phi_{\rm end}}\frac{V_{\phi}}{\chi_{\phi} V}d\phi\longrightarrow {\cal N}\equiv N+N_{\rm end}=\frac{1}{16 \lambda \sigma^2_{\star}}+\frac{\sigma^2_{\star}}{16}-\Bigg(\frac{\sigma^4_{\star}}{64}+\frac{\log \sigma_{\star}}{16 \lambda }\Bigg)\xi\,,\label{numn}
\end{eqnarray}
where $N_{\rm end}$ is derived
\begin{eqnarray}
N_{\rm end}=\frac{1}{16 \lambda \sigma^2_{\rm end}}+\frac{\sigma^2_{\rm end}}{16}-\Bigg(\frac{\sigma^4_{\rm end}}{64}+\frac{\log \sigma_{\rm end}}{16 \lambda }\Bigg)\xi\,.\label{numn0}
\end{eqnarray}
We next consider the above expression (\ref{numn}) and use a perturbation method in the small parameters $\xi$ and investigate
for a solution to this condition of the type:
\begin{eqnarray}
{\bar \sigma}_{\star}={\bar \sigma}_{0}+{\bar \sigma}_{1}\xi, \quad{\rm with}\quad {\bar \sigma}_{0}= \sqrt{8\,{\cal N}\,\Sigma}\,.
\end{eqnarray}
From (\ref{numn}), the solution reads
\begin{eqnarray}
{\bar \sigma}_{\star}&=&{\bar \sigma}_{0}+ \left[-\frac{\lambda {\bar \sigma}_{0}^7}{8 \left(1-\lambda  {\bar \sigma}_{0}^4\right)}+\frac{{\bar \sigma}_{0}^3 \log ({\bar \sigma}_{0})}{2 \left(1-\lambda  {\bar \sigma}_{0}^4\right)}\right]\xi +{\cal O}(\xi^{2}).
\end{eqnarray}
Taking the value of ${\bar \sigma}_{\star}$ at horizon crossing, the slow-roll parameters $\epsilon$ and $\eta$ can be conveniently rewritten as functions of the e-folding number $\cal N$ to first order in $\xi$:
\begin{eqnarray}
\epsilon_{\star}&=&\frac{4}{N}\frac{\Pi ^2 \Sigma ^3}{\left(1-\Pi  \Sigma ^2\right)^2}\nonumber\\&&+\frac{8 \Pi ^2 \Sigma^4}{\left(1-\Pi  \Sigma ^2\right)^3}\Bigg(4 \left(\Pi  \Sigma ^2+3\right) \log \left(\sqrt{8 M \Sigma }\right)+\Pi  \Sigma ^2 \left(3 \Pi  \Sigma ^2-11\right)+4\Bigg)\xi+{\cal O}(\xi^{2})\,,
\end{eqnarray}
and
\begin{eqnarray}
\eta_{\star}&=&-\frac{3}{N}\frac{\Pi  \Sigma }{1-\Pi  \Sigma ^2}+\frac{4}{N}\frac{\Pi ^2 \Sigma ^3}{\left(1-\Pi  \Sigma ^2\right)^2}\nonumber\\&&+\Bigg(\frac{24 \Pi ^3 \Sigma ^6}{\Pi  \Sigma ^2-1}+18 \Pi ^2 \Sigma ^4-\frac{88 \Pi ^2 \Sigma ^4}{\Pi  \Sigma ^2-1}-54 \Pi  \Sigma ^2+\frac{32 \Pi  \Sigma ^2}{\Pi  \Sigma ^2-1}+24\Bigg)\frac{\xi  \Pi  \Sigma ^2}{\left(\Pi  \Sigma ^2-1\right)^3}\nonumber\\&&+\Bigg(\frac{32 \Pi ^3 \Sigma ^6}{1-\Pi  \Sigma ^2}-24 \Pi ^2 \Sigma ^4+\frac{96 \Pi ^2 \Sigma ^4}{1-\Pi  \Sigma ^2}-24 \Pi  \Sigma ^2\Bigg)\frac{\xi  \log \left(\sqrt{8 M \Sigma }\right)}{\left(1-\Pi  \Sigma ^2\right)^3}+{\cal O}(\xi^{2}).
\end{eqnarray}
Using the above solution for ${\bar \sigma}={\bar \sigma}_{\star}$, the scalar spectral index $n_s$ and the tensor-to-scalar ratio $r$ can then be expressed to first order in $\xi$. This can be simply done by $n_s=1-6\epsilon_{\star}+2\eta_{\star}$ and $r=16\epsilon_{\star}$. However, it can be straightforwardly verified that, in the limit $\xi \rightarrow 0$, the results smoothly reduce to those of the minimally coupled model, confirming the consistency of our formulation with the minimal case. 
\begin{table}[t]
\centering
	\begin{tabular}{c|c|c|c|c|c|c|c|c|c}
	\quad ${\cal N}$ \quad\quad & \quad$ \lambda$ \quad\quad& \quad $\xi$ \quad\quad & \quad\,\, $r$ \quad\quad & \quad$n_s$\quad\quad & \quad ${\cal N}$ \quad\quad & \quad\quad $\lambda$ \quad\quad & \quad\quad $\xi$ \quad\quad & \quad $r$ \quad\quad & \quad $n_s$ \quad\quad \\\hline
  80  &       &       &  $0.00260$  &  $0.9607$   &  80   &      &      &   $0.000245$   &  $0.9624$  \\
 90   &     &      &  $0.00181$  &  $0.9653$   &   90  &      &      &  $0.000172$ &  $0.9666$  \\
    $100$ &  $10^{-4}$  & $10^{-3}$ & $0.00131$ & $0.9690$  & $117$ & $10^{-3}$ & $10^{-3}$ & $0.0000781$ & $0.9743$\\
 110   &     &     & $0.000974$ &  $0.9720$   &  118 &       &     & $0.0000762$ & $0.9745$\\
 120   &      &    & $0.0067$ &  $0.9791$  &  120  &       &        & $0.000747$ & $0.9744$\\\hline
	\end{tabular}
	\caption{We present specific predictions for the tensor-to-scalar ratio $r$ and the scalar spectral index $n_s$ in the non-minimally coupled model for a weakly coupled case ($\xi \ll 1$), fixing ${\cal N}=100$ and ${\cal N}=120$, while varying $\lambda$.} \label{tab:non1}
\end{table}
Considering the amplitude of scalar perturbations given in Eq.(\ref{10}), $V_{0}/M^{4}_{p}$ can be directly determined to obtain:
\begin{eqnarray}
\frac{V_{0}}{M^{4}_{p}} &\simeq& \frac{1.59\times 10^{-5} \lambda ^2 {\bar \sigma}_{\star}^6}{\left(1-\lambda {\bar \sigma}_{\star}^4\right)^2 \left(1-3 \xi {\bar \sigma}_{\star}^2\right) \left(1-\lambda  {\bar \sigma}_{\star}^4\right)^2}\,.\label{beta}
\end{eqnarray}
In the limit $\xi \to 0$, the results smoothly approach those of the minimally-coupled case. For nonminimally-coupled model, we particularly report the predictions for the tensor-to-scalar ratio $r$ and the spectral index $n_s$ with $\lambda=10^{-4},\,10^{-3}$ and $\xi=10^{-3}$ for various values of ${\cal N}$ in Table \ref{tab:non1}. We find that the nonminimally-coupled model with weakly coupled case using $\lambda=10^{-3},\,\xi=10^{-3}$ and ${\cal N}=117$ predicts the tensor-to-scalar ratio $r\sim 7.8\times 10^{-5}$ and the spectral index $n_s=0.9743$ in excellent agreement with those reported by the observations. However, matching the model predictions with the observations demands unusually large
values of the effective e-folding number, $\mathcal{N}$, which may be questionable compared with the conventional range ($N\simeq55$-$60$).

\subsubsection*{2.2) Strongly coupled case ($\xi \gg 1$)}
We start by introducing the dimensionless variable ${\tilde \psi} \equiv \phi/M_{p}$ such that the slow-roll parameters are expressed as follows
\begin{eqnarray}
\epsilon \simeq \frac{16 \lambda ^2 {\tilde \psi}^8}{3 \left(1-\lambda {\tilde \psi}^4\right)^2},
\qquad
\eta \simeq \frac{16 \lambda {\tilde \psi}^4 \left(2 \lambda  {\tilde \psi}^4-1\right)}{3 \left(\lambda {\tilde \psi}^4-1\right)^2}.
\label{13zl}
\end{eqnarray}
The value of the field at the end of inflation, ${\tilde \psi}_{\text{end}}$, is determined by the condition $\epsilon = 1$, yielding
\begin{gather}
\frac{16\lambda^2 {\tilde \psi}^8_{\text{end}}}{3\left(1-\lambda  {\tilde \psi}^4_{\text{end}}\right)^2} = 1 \longrightarrow  {\tilde \psi}_{\text{end}}=\Bigg(\frac{4 \sqrt{3}}{13 \lambda }-\frac{3}{13 \lambda }\Bigg)^{-1/4},
\label{14z}
\end{gather}
Integrating the number of $e$-folds, we obtain
\begin{gather}
N = \frac{1}{M_{p}^{2}} \int_{\phi_{\text{end}}}^{\phi_{\star}} \frac{\chi_{\phi} V }{V_{\phi}}\,d\phi\quad\longrightarrow\quad {\cal N}=N+N_{\rm end}=\frac{\sqrt{\frac{3}{2}}}{4 \lambda }\left(\frac{1}{3 {\tilde \psi}^3}+\lambda  {\tilde \psi}\right),
\end{gather}
Therefore, the field value at horizon crossing, ${\tilde \psi}={\tilde \psi}_{\star}$, can be obtained as
\begin{eqnarray}
{\tilde \psi}_{\star} &=& \sqrt{\frac{2}{3}} {\cal N}+\frac{1}{\sqrt{6}}\sqrt{\Gamma +\frac{1}{\Gamma }+4 {\cal N}^2}-\frac{1}{2} \sqrt{-\frac{2 \Gamma }{3}-\frac{2}{3 \Gamma }+\frac{16 {\cal N}^2}{3}+\frac{32 {\cal N}^3}{3 \sqrt{\Gamma +\frac{1}{\Gamma }+4 {\cal N}^2}}}\,,
\label{15zzl}
\end{eqnarray}
where we have defined a new parameter $\Gamma$ as:
\begin{eqnarray}
\Gamma=\lambda\Bigg(\frac{\sqrt{36 \lambda  {\cal N}^4-1}}{\lambda ^{3/2}}+\frac{6 {\cal N}^2}{\lambda }\Bigg)^{1/3}\,.
\end{eqnarray}
In the above, we have defined
\begin{eqnarray}
N_{\rm end}=\frac{\sqrt{\frac{3}{2}}}{4 \lambda }\left(\frac{1}{3 {\tilde \psi}^3_{\rm end}}+\lambda  {\tilde \psi}_{\rm end}\right)\,,
\label{15z}
\end{eqnarray}
where ${\tilde \psi}^{2}_{\rm end}$ is given in Eq.(\ref{14}). The principal inflationary observables, the scalar spectral index $n_{s}$ and the tensor-to-scalar ratio $r$, are expressed as
\begin{eqnarray}
n_{s} &=& 1 - 6\epsilon + 2\eta 
\nonumber\\
&=& 1-\frac{32 \lambda^2 {\tilde \psi}^8_{\star}}{\left(1-\lambda {\tilde \psi}^4_{\star}\right)^2}+\frac{32 \lambda {\tilde \psi}^4 \left(2 \lambda {\tilde \psi}^4_{\star}-1\right)}{3 \left(\lambda {\tilde \psi}^4_{\star}-1\right)^2}, \label{16zll}
\\
r &=& 16\epsilon = \frac{256 \lambda^2 {\tilde \psi}^8_{\star}}{3 \left(1-\lambda {\tilde \psi}^4_{\star}\right)^2}.
\label{16zl2}
\end{eqnarray}
The above parameters can be simply recast in terms of the $e$-folding number ${\cal N}$ by substituting ${\tilde \psi}_{\star}$ given in Eq.(\ref{15zzl}) to Eqs.(\ref{16zll}) and (\ref{16zl2}).
For this strongly coupled scenario, we particularly present the predictions for the tensor-to-scalar ratio $r$ and spectral index $n_{s}$ with ${\cal N} = 65$ and ${\cal N} = 70$ for representative values of $\lambda$ in Table~\ref{tab:non2}. Considering the amplitude of scalar perturbations given in Eq.(\ref{10}), $V_{0}/M^{4}_{p}$ can be directly determined to obtain:
\begin{eqnarray}
\frac{V_{0}}{M^{4}_{p}} &\simeq& \frac{2.65\times 10^{-6} \lambda^2 {\tilde \psi}^8_{\star} \left(1+\xi {\tilde \psi}^2_{\star}\right)^2}{\left(1-\lambda {\tilde \psi}^4_{\star}\right)^4}\,,\label{betal}
\end{eqnarray}
where ${\tilde \psi}_{\star}$ given in Eq.(\ref{15zzl}). It is found that the characteristic energy scale $V^{1/4}_{0}$ is found to be of the order of $[10^{-3}-10^{-2}]\,M_{p}$.
\begin{table}[t]
\centering
	\begin{tabular}{c|c|c|c|c|c|c|c}
	\quad ${\cal N}$ \quad\quad & \quad $\lambda$ \quad\quad & \quad\,\, $r$ \quad\quad & \quad$n_s$\quad\quad & \quad ${\cal N}$ \quad\quad &\quad $\lambda$ \quad\quad & \quad $r$ \quad\quad & \quad $n_s$ \quad\quad  \\\hline
    &  0.000214  & 0.000819  &  $0.9667$   &     & 0.00014 & 0.000893  & 0.9653     \\
    &  0.000325  &  0.000617 &  $0.9712$   &     & 0.00024 & 0.000620  &   0.9711   \\\hline
$65$&  0.000456  &  0.000491 & $0.9743$    & $70$& 0.00034 & 0.000490 & 0.9743 \\\hline
    &  0.000502  &  0.000460 &  $0.9751$   &     & 0.00054  & 0.000359  & 0.9780   \\
    &  0.000635  &  0.000393 &  $0.9770$   &     &  0.00080 & 0.000276  & 0.9807    \\\hline
	\end{tabular}
	\caption{We present specific predictions for the tensor-to-scalar ratio $r$ and the scalar spectral index $n_s$ in the non-minimally coupled model for a strongly coupled case ($\xi \gg 1$), fixing ${\cal N}=65$ and ${\cal N}=70$, while varying $\lambda$.} \label{tab:non2}
\end{table}

\section{Confrontation with ACT Constraints}\label{III}
Now we compare the predictions of the tensor-to-scalar ratio $r$ and the scalar spectral index $n_s$ with the ACT data \cite{ACT:2025fju,ACT:2025tim}, the Planck data \cite{Planck:2018jri} and the updated Planck constraints on the tensor-to-scalar ratio \cite{BICEP:2021xfz}. We start with the minimal case, and then the non-minimally-coupled model.
\subsubsection*{3.1) Weakly coupled case ($\xi \ll 1$)}
\begin{figure}
\includegraphics[width=7.5 cm]{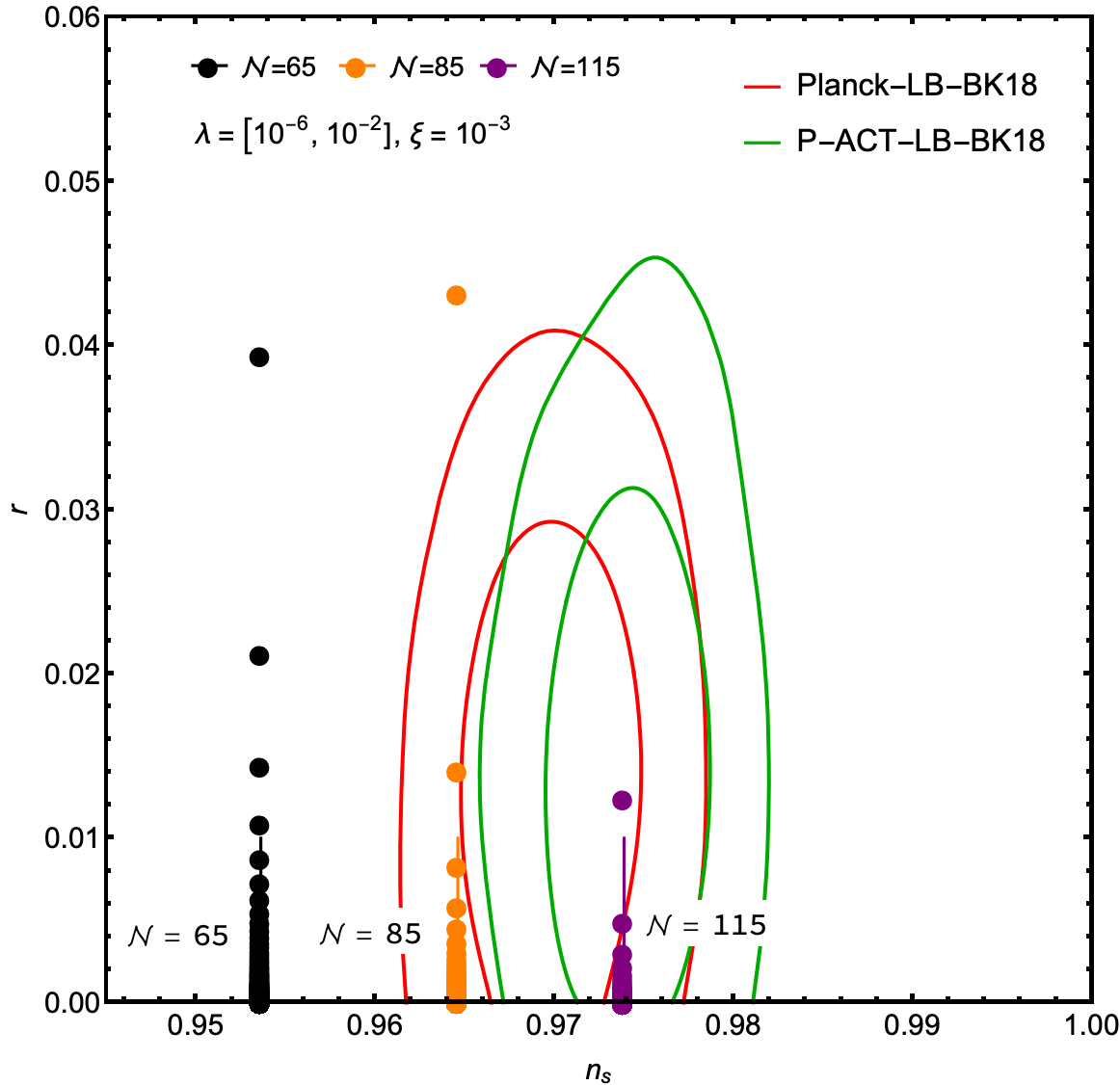}
\includegraphics[width=7.5 cm]{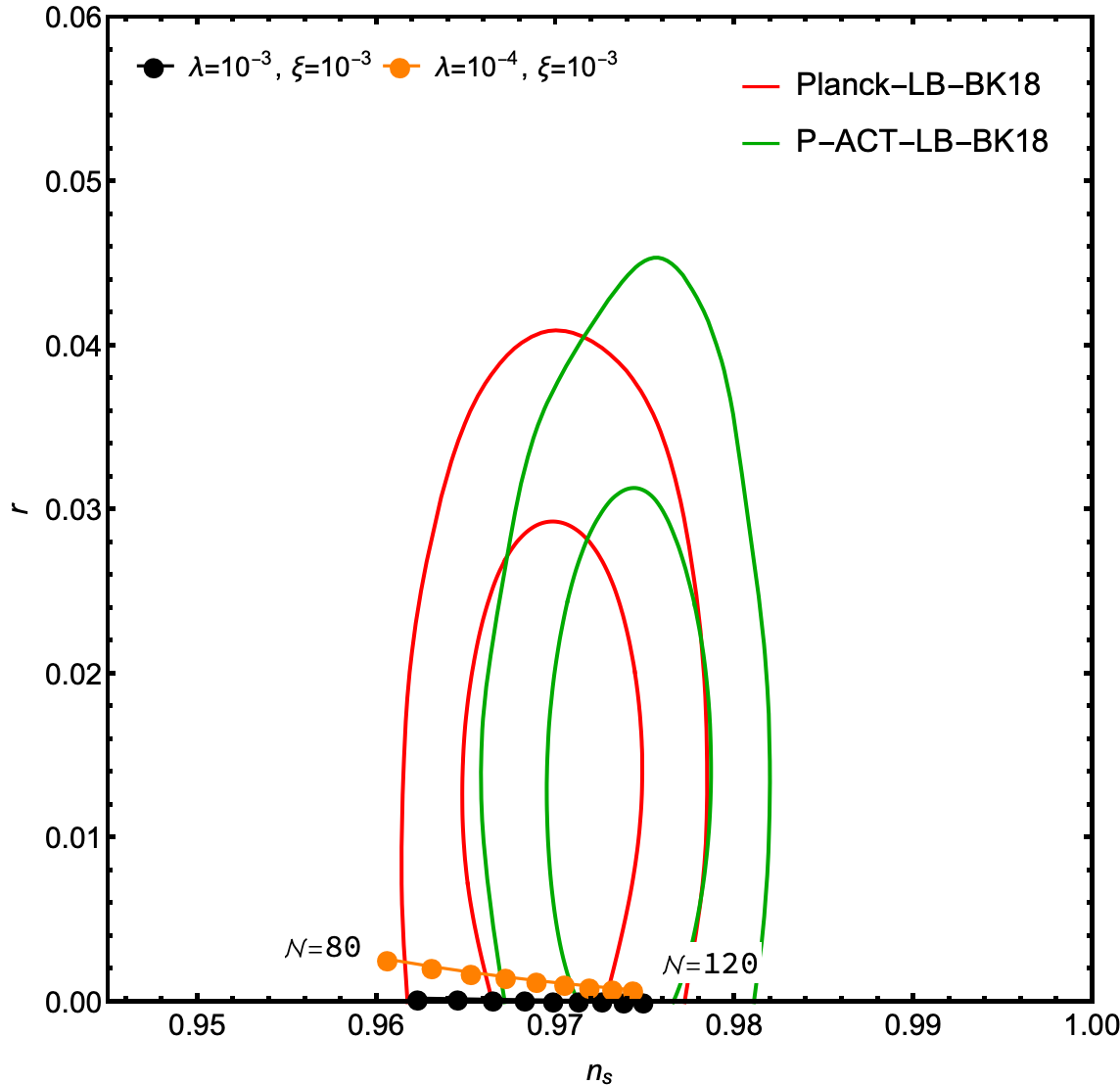}
\caption{Constraints on the scalar and tensor primordial power spectra, shown in the $r-n_{s}$
parameter space predicted by the nonminimally-coupled model for a weakly coupled case ($\xi \ll 1$). Left panel: ${\cal N}=[65,\,85,\,115]$ are kept fixed, while varying $\lambda$; Right panel:  we choose $\lambda=10^{-3}$ and $\lambda=10^{-4}$ and fix a parameter $\xi=10^{-3}$, while varying $\cal N$. The bounds on $r$ are primarily determined by the BK18 observations, whereas the limits on $n_s$ are set by Planck (red) and P-ACT (green) data.}\label{rnsnon}
\end{figure}

In the weakly coupled regime ($\xi\ll1$), Fig.~\ref{rnsnon} shows that increasing the e-fold number drives the model predictions toward smaller $r$ and larger $n_s$, improving agreement with the
ACT/Planck contours; this is also reflected in our weak-coupling benchmarks reported at
${\cal N}=100$ and ${\cal N}=120$, See Fig.~\ref{rnsnon} for the weak-coupling scans and Table~\ref{rnsnon} for explicit
entries at ${\cal N}=100,\,120$. This trend follows from the slow-roll integrals: a longer inflationary
phase effectively flattens the trajectory in the Einstein frame and suppresses tensors.

\subsubsection*{3.2) Strongly coupled case ($\xi \gg 1$)}
\begin{figure}
\includegraphics[width=10 cm]{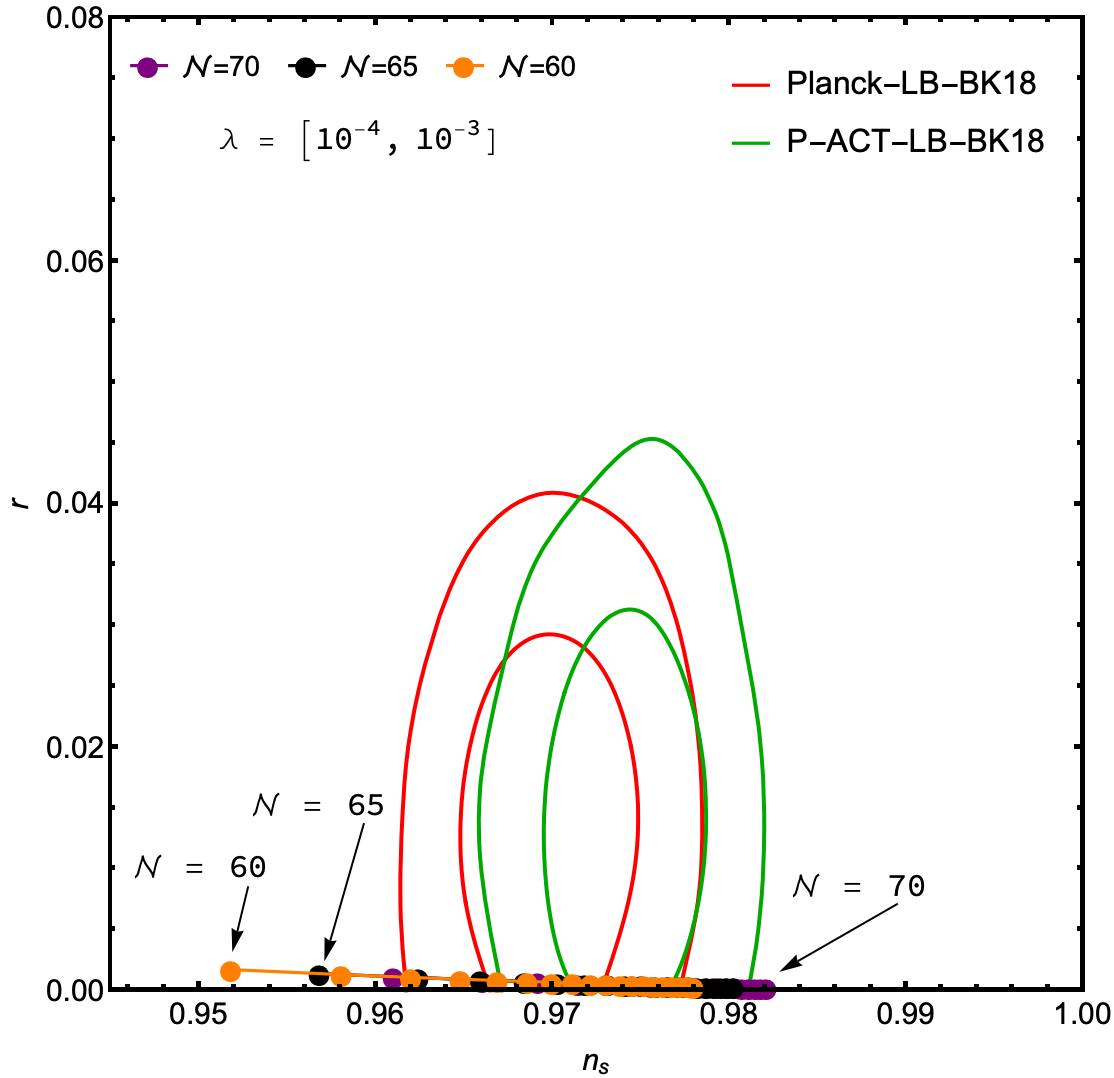}
\caption{Constraints on the scalar and tensor primordial power spectra, shown in the $r-n_{s}$
parameter space predicted by the nonminimally-coupled model for a strongly coupled case ($\xi \gg 1$). ${\cal N}=[60,\,65,\,70]$ are kept fixed, while varying $\lambda=[10^{-4},\,10^{-3}]$. The bounds on $r$ are primarily determined by the BK18 observations, whereas the limits on $n_s$ are set by Planck (red) and P-ACT (green) data.}\label{rnsnonl}
\end{figure}

In the strongly coupled case, conformal rescaling yields an exponentially flat plateau, so competitive $(n_s,r)$ is already achieved for ${\cal N}\simeq 65\!-\!70$. Our benchmarks (Table~\ref{rnsnonl}) give $n_s\simeq0.9743$ with $r\lesssim5\times10^{-4}$ in this range, consistent with the observational contours (see Fig.~\ref{rnsnonl}). The corresponding
  normalization remains viable with $V_0^{1/4}\sim 10^{-3}\!-\!10^{-2}M_p$.

\section{Conclusions}\label{Con}
In this work, we have investigated the inflationary dynamics of the squared--Quartic Hilltop potential in both minimally and non-minimally coupled frameworks, motivated by the recent Atacama Cosmology Telescope (ACT) results that report a higher scalar spectral index $n_s$ than the Planck 2018 value. Beginning with the canonical minimally coupled case, we derived analytic expressions for the slow-roll parameters and the corresponding observables $n_s$ and $r$, showing that the model can reproduce viable predictions within the Planck confidence limits.

Extending the analysis to include a non-minimal coupling between the scalar field and gravity, we examined two distinct regimes of the coupling parameter $\xi$. In the weakly coupled limit ($\xi \ll 1$), we obtained perturbative corrections to the standard slow-roll expressions and demonstrated that the inclusion of a small positive coupling slightly increases $n_s$ while suppressing $r$. The model predicts $n_s \simeq 0.9743$ and $r \sim 7.8\times10^{-5}$ for representative parameters ($\lambda = 10^{-3}$, $\xi = 10^{-3}$, ${\cal N} = 117$), in excellent agreement with the joint Planck-ACT-DESI (P-ACT-LB) constraints. However, this regime requires a relatively large effective number of e-folds to reach full consistency with observations. In the strongly coupled regime ($\xi \gg 1$), the conformal rescaling induces a pronounced flattening of the potential, leading to an exponentially suppressed tensor-to-scalar ratio. The predictions converge to $n_s \approx 0.9743$ with $r \lesssim 5\times10^{-4}$ for ${\cal N} \simeq 65-70$, which lie well within the 68\% confidence region of ACT and BK18 bounds. The associated energy scale of inflation, $V_0^{1/4} \sim 10^{-3}-10^{-2}M_p$, remains consistent with high-scale inflationary models.

Our analysis shows that the non-minimally coupled squared--Quartic Hilltop inflation provides a robust and flexible framework that can successfully accommodate the new ACT results. The model naturally interpolates between the minimally coupled limit and the strongly flattened plateau regime, maintaining theoretical consistency while yielding predictions fully compatible with the most recent cosmological data. Future high-precision CMB polarization experiments are expected to further test the small tensor predictions and constrain the viable range of the coupling parameter $\xi$. Future investigations may focus on extending the present framework to incorporate radiative and finite-temperature corrections, as well as on examining the reheating dynamics and their implications for key inflationary observables. Another promising direction involves testing the model’s compatibility with swampland conjectures and related consistency conditions. Upcoming high-precision CMB missions, such as CORE~\cite{COrE:2011bfs}, AliCPT~\cite{Li:2017drr}, LiteBIRD~\cite{Matsumura:2013aja}, and CMB-S4~\cite{Abazajian:2019eic}, are expected to provide critical tests for these low-$r$ inflationary frameworks.

\section*{Acknowledgement}\label{ac}
This work has received funding support from
the NSRF via the Program Management Unit for Human Resources \& Institutional Development, Research and Innovation [grant number B39G680009]



\begin{thebibliography}{99}
\bibitem{ACT:2025fju}
T.~Louis \textit{et al.} [ACT],
[arXiv:2503.14452 [astro-ph.CO]].

\bibitem{ACT:2025tim}
E.~Calabrese \textit{et al.} [ACT],
[arXiv:2503.14454 [astro-ph.CO]].

\bibitem{DESI:2024uvr}
A.~G.~Adame \textit{et al.} [DESI],
JCAP \textbf{04} (2025), 012 

\bibitem{DESI:2024mwx}
A.~G.~Adame \textit{et al.} [DESI],
JCAP \textbf{02} (2025), 021

\bibitem{Planck:2018jri}
Y.~Akrami \textit{et al.} [Planck],
Astron. Astrophys. \textbf{641} (2020), A10

\bibitem{Starobinsky:1980te}
A.~A.~Starobinsky,
Phys. Lett. B \textbf{91} (1980), 99-102

\bibitem{Sato:1981qmu}
K.~Sato,
Mon. Not. Roy. Astron. Soc. \textbf{195} (1981) no.3, 467-479

\bibitem{Guth:1980zm}
A.~H.~Guth,
Phys. Rev. D \textbf{23} (1981), 347-356

\bibitem{Linde:1981mu}
A.~D.~Linde,
Phys. Lett. B \textbf{108} (1982), 389-393

\bibitem{Albrecht:1982wi}
A.~Albrecht and P.~J.~Steinhardt,
Phys. Rev. Lett. \textbf{48} (1982), 1220-1223

\bibitem{Kallosh:2013tua}
R.~Kallosh, A.~Linde and D.~Roest,
Phys. Rev. Lett. \textbf{112} (2014) no.1, 011303

\bibitem{Kallosh:2013hoa}
R.~Kallosh and A.~Linde,
JCAP \textbf{07} (2013), 002

\bibitem{Kallosh:2013maa}
R.~Kallosh and A.~Linde,
JCAP \textbf{10} (2013), 033

\bibitem{Kallosh:2013yoa}
R.~Kallosh, A.~Linde and D.~Roest,
JHEP \textbf{11} (2013), 198

\bibitem{Kallosh:2014rga}
R.~Kallosh, A.~Linde and D.~Roest,
JHEP \textbf{08} (2014), 052

\bibitem{Kallosh:2015lwa}
R.~Kallosh and A.~Linde,
Phys. Rev. D \textbf{91} (2015), 083528

\bibitem{Roest:2015qya}
D.~Roest and M.~Scalisi,
Phys. Rev. D \textbf{92} (2015), 043525

\bibitem{Linde:2016uec}
A.~Linde,
JCAP \textbf{02} (2017), 028

\bibitem{Terada:2016nqg}
T.~Terada,
Phys. Lett. B \textbf{760} (2016), 674-680

\bibitem{Ueno:2016dim}
Y.~Ueno and K.~Yamamoto,
Phys. Rev. D \textbf{93} (2016) no.8, 083524

\bibitem{Odintsov:2016vzz}
S.~D.~Odintsov and V.~K.~Oikonomou,
Phys. Rev. D \textbf{94} (2016) no.12, 124026

\bibitem{Akrami:2017cir}
Y.~Akrami, R.~Kallosh, A.~Linde and V.~Vardanyan,
JCAP \textbf{06} (2018), 041

\bibitem{Dimopoulos:2017zvq}
K.~Dimopoulos and C.~Owen,
JCAP \textbf{06} (2017), 027

\bibitem{Barrow:2016qkh}
J.~D.~Barrow and A.~Paliathanasis,
Phys. Rev. D \textbf{94} (2016) no.8, 083518
doi:10.1103/PhysRevD.94.083518
[arXiv:1609.01126 [gr-qc]].


\bibitem{Kaiser:1994vs}
D.~I.~Kaiser,
Phys. Rev. D \textbf{52} (1995), 4295-4306

\bibitem{Bezrukov:2008ej}
F.~L.~Bezrukov, A.~Magnin and M.~Shaposhnikov,
Phys. Lett. B \textbf{675} (2009), 88-92

\bibitem{Bezrukov:2007ep}
F.~L.~Bezrukov and M.~Shaposhnikov,
Phys. Lett. B \textbf{659} (2008), 703-706

\bibitem{Chang:2019ebx}
P.~W.~Chang, C.~W.~Chiang and K.~W.~Ng,
JHEP \textbf{04} (2020), 163


\bibitem{Karwan:2013iph}
K.~Karwan and P.~Channuie,
JCAP \textbf{06} (2014), 045

\bibitem{Channuie:2012bv}
P.~Channuie, J.~J.~Jorgensen and F.~Sannino,
Phys. Rev. D \textbf{86} (2012), 125035

\bibitem{Bezrukov:2011mv}
F.~Bezrukov, P.~Channuie, J.~J.~Joergensen and F.~Sannino,
Phys. Rev. D \textbf{86} (2012), 063513

\bibitem{Channuie:2011rq}
P.~Channuie, J.~J.~Joergensen and F.~Sannino,
JCAP \textbf{05} (2011), 007

\bibitem{Channuie:2014ysa}
P.~Channuie,
Nucl. Phys. B \textbf{892} (2015), 429-448

\bibitem{Samart:2022pza}
D.~Samart, C.~Pongkitivanichkul and P.~Channuie,
Eur. Phys. J. ST \textbf{231} (2022) no.7, 1325-1344

\bibitem{Planck:2018jri}
Y.~Akrami \textit{et al.} [Planck],
Astron. Astrophys. \textbf{641} (2020), A10

\bibitem{Gialamas:2025kef}
I.~D.~Gialamas, A.~Karam, A.~Racioppi and M.~Raidal,
[arXiv:2504.06002 [astro-ph.CO]].

\bibitem{Kallosh:2025rni}
R.~Kallosh, A.~Linde and D.~Roest,
[arXiv:2503.21030 [hep-th]].


\bibitem{Gao:2025onc}
Q.~Gao, Y.~Gong, Z.~Yi and F.~Zhang,
[arXiv:2504.15218 [astro-ph.CO]].



\bibitem{Liu:2025qca}
L.~Liu, Z.~Yi and Y.~Gong,
[arXiv:2505.02407 [astro-ph.CO]].

\bibitem{Yogesh:2025wak}
Yogesh, A.~Mohammadi, Q.~Wu and T.~Zhu,
[arXiv:2505.05363 [astro-ph.CO]].

\bibitem{Yi:2025dms}
Z.~Yi, X.~Wang, Q.~Gao and Y.~Gong,
[arXiv:2505.10268 [astro-ph.CO]].


\bibitem{Peng:2025bws}
Z.~Z.~Peng, Z.~C.~Chen and L.~Liu,
[arXiv:2505.12816 [astro-ph.CO]].

\bibitem{Yin:2025rrs}
W.~Yin,
[arXiv:2505.03004 [hep-ph]].

\bibitem{Byrnes:2025kit}
C.~T.~Byrnes, M.~Cort\^es and A.~R.~Liddle,
[arXiv:2505.09682 [astro-ph.CO]].

\bibitem{Wolf:2025ecy}
W.~J.~Wolf,
[arXiv:2506.12436 [astro-ph.CO]].

\bibitem{Aoki:2025wld}
S.~Aoki, H.~Otsuka and R.~Yanagita,
[arXiv:2504.01622 [hep-ph]].


\bibitem{Gao:2025viy}
Q.~Gao, Y.~Qian, Y.~Gong and Z.~Yi,
[arXiv:2506.18456 [gr-qc]].



\bibitem{Zahoor:2025nuq}
M.~Zahoor, S.~Khan and I.~A.~Bhat,
[arXiv:2507.18684 [astro-ph.CO]].


\bibitem{Ferreira:2025lrd}
E.~G.~M.~Ferreira, E.~McDonough, L.~Balkenhol, R.~Kallosh, L.~Knox
and A.~Linde,
[arXiv:2507.12459 [astro-ph.CO]].


\bibitem{Mohammadi:2025gbu}
A.~Mohammadi, Yogesh and A.~Wang,
[arXiv:2507.06544 [astro-ph.CO]].

\bibitem{Choudhury:2025vso}
S.~Choudhury, G.~Bauyrzhan, S.~K.~Singh and K.~Yerzhanov,
[arXiv:2506.15407 [astro-ph.CO]].

\bibitem{Odintsov:2025eiv}
S.~D.~Odintsov and V.~K.~Oikonomou,
Phys. Lett. B \textbf{870} (2025), 139907

\bibitem{Odintsov:2025wai}
S.~D.~Odintsov and V.~K.~Oikonomou,
Phys. Lett. B \textbf{868} (2025), 139779


\bibitem{Q:2025ycf}
R.~D.~A.~Q., J.~Chagoya and A.~A.~Roque,
[arXiv:2508.13273 [gr-qc]].


\bibitem{Zhu:2025twm}
Y.~Zhu, Q.~Gao, Y.~Gong and Z.~Yi,
[arXiv:2508.09707 [astro-ph.CO]].

\bibitem{Kouniatalis:2025orn}
G.~Kouniatalis and E.~N.~Saridakis,
[arXiv:2507.17721 [astro-ph.CO]].


\bibitem{Hai:2025wvs}
M.~Hai, A.~R.~Kamal, N.~F.~Shamma and M.~S.~J.~Shuvo,
[arXiv:2506.08083 [hep-th]].


\bibitem{Dioguardi:2025vci}
C.~Dioguardi, A.~J.~Iovino and A.~Racioppi,
Phys. Lett. B \textbf{868} (2025), 139664

\bibitem{Yuennan:2025kde}
J.~Yuennan, P.~Koad, F.~Atamurotov and P.~Channuie,
Eur. Phys. J. C \textbf{85} (2025) no.11, 1307

\bibitem{Oikonomou:2025xms}
V.~K.~Oikonomou,
[arXiv:2508.19196 [gr-qc]].

\bibitem{Oikonomou:2025htz}
V.~K.~Oikonomou,
Phys. Lett. B \textbf{871} (2025), 139972

\bibitem{Odintsov:2025jky}
S.~D.~Odintsov and V.~K.~Oikonomou,
Phys. Lett. B \textbf{870} (2025), 139909

\bibitem{Aoki:2025ywt}
S.~Aoki, H.~Otsuka and R.~Yanagita,
[arXiv:2509.06739 [hep-ph]].

\bibitem{Gialamas:2025ofz}
I.~D.~Gialamas, T.~Katsoulas and K.~Tamvakis,
JCAP \textbf{09} (2025), 060

\bibitem{Yuennan:2025tyx}
J.~Yuennan, F.~Atamurotov and P.~Channuie,
[arXiv:2509.23329 [gr-qc]].

\bibitem{Pallis:2025nrv}
C.~Pallis,
Phys. Lett. B \textbf{868} (2025), 139739

\bibitem{Pallis:2025gii}
C.~Pallis,
JCAP \textbf{09} (2025), 061

\bibitem{Yuennan:2025inm}
J.~Yuennan, F.~Atamurotov and P.~Channuie,
Phys. Lett. B \textbf{871} (2025), 139958

\bibitem{Lynker:2025wyc}
M.~Lynker and R.~Schimmrigk,
[arXiv:2507.15076 [astro-ph.CO]].

\bibitem{Kallosh:2025ijd}
R.~Kallosh and A.~Linde,
[arXiv:2505.13646 [hep-th]].

\bibitem{Hoffmann:2021vty}
J.~Hoffmann and D.~Sloan,
Phys. Rev. D \textbf{104} (2021) no.12, 123542


\bibitem{Dimopoulos:2020kol}
K.~Dimopoulos,
Phys. Lett. B \textbf{809} (2020), 135688
doi:10.1016/j.physletb.2020.135688
[arXiv:2006.06029 [hep-ph]].

\bibitem{German:2020rpn}
G.~German,
JCAP \textbf{02} (2021), 034

\bibitem{Schwinger:1951nm}
J.~S.~Schwinger,
Phys. Rev. \textbf{82} (1951), 664-679

\bibitem{DeWitt:1975ys}
B.~S.~DeWitt,
Phys. Rept. \textbf{19} (1975), 295-357

\bibitem{Callan:1970ze}
C.~G.~Callan, Jr., S.~R.~Coleman and R.~Jackiw,
Annals Phys. \textbf{59} (1970), 42-73

\bibitem{Buchbinder:1985ew}
I.~L.~Buchbinder,
Fortsch. Phys. \textbf{34} (1986), 605-628

\bibitem{Odintsov:1990mt}
S.~D.~Odintsov,
Fortsch. Phys. \textbf{39} (1991), 621-641
Print-90-0237 (TOMSK).

\bibitem{Muta:1991mw}
T.~Muta and S.~D.~Odintsov,
Mod. Phys. Lett. A \textbf{6} (1991), 3641-3646

\bibitem{Elizalde:1994im}
E.~Elizalde and S.~D.~Odintsov,
Phys. Lett. B \textbf{333} (1994), 331-336

\bibitem{Buchbinder:1989bt}
I.~L.~Buchbinder, S.~D.~Odintsov and I.~M.~Lichtzier,
Class. Quant. Grav. \textbf{6} (1989), 605-610

\bibitem{Chernikov:1968zm}
N.~A.~Chernikov and E.~A.~Tagirov,
Ann. Inst. H. Poincare Phys. Theor. A \textbf{9} (1968) no.2, 109-141

\bibitem{Parker:1983pe}
L.~Parker and D.~J.~Toms,
Phys. Rev. D \textbf{29} (1984), 1584

\bibitem{Boubekeur:2005zm}
L.~Boubekeur and D.~H.~Lyth,
JCAP \textbf{07} (2005), 010

\bibitem{Kinney:1995cc}
W.~H.~Kinney and K.~T.~Mahanthappa,
Phys. Rev. D \textbf{53} (1996), 5455-5467

\bibitem{Lillepalu:2022knx}
H.~G.~Lillepalu and A.~Racioppi,
Eur. Phys. J. Plus \textbf{138} (2023) no.10, 894

\bibitem{Lynker:2023hfm}
M.~Lynker and R.~Schimmrigk,
Phys. Rev. D \textbf{110} (2024) no.2, 023508

\bibitem{Kallosh:2019jnl}
R.~Kallosh and A.~Linde,
JCAP \textbf{09} (2019), 030

\bibitem{Bostan:2025may}
N.~Bostan and R.~H.~Dejrah,
Nucl. Phys. B \textbf{1018} (2025), 117056

\bibitem{Bostan:2024ugi}
N.~Bostan and R.~H.~Dejrah,
[arXiv:2409.10398 [astro-ph.CO]].

\bibitem{Lyth:1998xn}
D.~H.~Lyth and A.~Riotto,
Phys. Rept. \textbf{314} (1999), 1-146

\bibitem{BICEP:2021xfz}
P.~A.~R.~Ade \textit{et al.} [BICEP and Keck],
Phys. Rev. Lett. \textbf{127} (2021) no.15, 151301

\bibitem{COrE:2011bfs}
F.~R.~Bouchet \textit{et al.} [COrE],
[arXiv:1102.2181 [astro-ph.CO]].

\bibitem{Li:2017drr}
H.~Li, S.~Y.~Li, Y.~Liu, Y.~P.~Li, Y.~Cai, M.~Li, G.~B.~Zhao, C.~Z.~Liu, Z.~W.~Li and H.~Xu, \textit{et al.}
Natl. Sci. Rev. \textbf{6} (2019) no.1, 145-154

\bibitem{Matsumura:2013aja}
T.~Matsumura, Y.~Akiba, J.~Borrill, Y.~Chinone, M.~Dobbs, H.~Fuke, A.~Ghribi, M.~Hasegawa, K.~Hattori and M.~Hattori, \textit{et al.}
J. Low Temp. Phys. \textbf{176} (2014), 733

\bibitem{Abazajian:2019eic}
K.~Abazajian, G.~Addison, P.~Adshead, Z.~Ahmed, S.~W.~Allen, D.~Alonso, M.~Alvarez, A.~Anderson, K.~S.~Arnold and C.~Baccigalupi, \textit{et al.}
[arXiv:1907.04473 [astro-ph.IM]].

\end{thebibliography}
\end{document}